\documentclass[twocolumn]{aastex6}
\usepackage{graphicx}
\shorttitle{HI of lensed galaxies}
\shortauthors{Edel, Hunt, Pisano}
\begin{document}
\title{The Search for HI Emission at z $\approx0.4$ in Gravitationally Lensed Galaxies with the Green Bank Telescope}
\author{L.~R.~Hunt\altaffilmark{1}, D.~J.~Pisano\altaffilmark{1}, S.~Edel\altaffilmark{2}}
\altaffiltext{1}{Department of Physics \& Astronomy, West Virginia University, P.O. Box 6315, Morgantown, WV 26506, USA, E-mail:lhunt3@mix.wvu.edu,djpisano@mail.wvu.edu}
\altaffiltext{2}{Infinite Optics,1712 Newport Cir \# F, Santa Ana, CA 92705, USA, E-mail:stasedel@gmail.com}
\begin{abstract}
Neutral Hydrogen (HI) provides a very important fuel for star formation, but is difficult to detect at high redshift due to weak emission, limited sensitivity of modern instruments, and terrestrial radio frequency interference (RFI) at low frequencies. We the first attempt to use gravitational lensing to detect HI line emission from three gravitationally lensed galaxies behind the cluster Abell 773, two at redshift of $0.398$ and one at $z=0.487$, using the Green Bank Telescope. We find a $3\sigma$ upper limit for a galaxy with a rotation velocity of 200 km s$^{-1}$ is M$_{HI}=6.58$$\times10^{9}$ and 1.5$\times$ $10^{10}$ $M_{\odot}$ at $z=0.398$ and $z=0.487$.  The estimated HI masses of the sources at z$=0.398$ and z$=0.487$ are a factor of $3.7$ and $\sim30$ times lower than our detection limits at the respective redshifts. To facilitate these observations we have used sigma clipping to remove both narrow- and wide-band RFI but retain the signal from the source. We are able to reduce the noise of the spectrum by $\sim25\%$ using our routine instead of discarding observations with too much RFI. The routine is most effective when $\sim10\%$ of the integrations or fewer contains RFI. These techniques can be used to study HI in highly magnified distant galaxies that are otherwise too faint to detect.
\end{abstract}
\section{Introduction}
\label{intro}
In order to better understand the star formation history of the Universe we must study the three indicators of star formation and its sustainability as a function redshift: changes of stellar mass, star formation rate and gas content of galaxies. Studies of stellar mass density between redshifts, z, of zero and four have shown that approximately $10\%$ of today's stellar mass has formed by $z\approx3$ and $50\%$ to $75\%$ has formed at $z\approx1$ \citep{2003ApJ...587...25D,2003ApJ...599..847R,2003ApJ...594L...9F} with the total stellar mass density increasing by an order of magnitude between $z=3.5$ and $z=0.1$ \citep{2013A&A...556A..55I,2009ApJ...701.1765M,2008ApJ...675..234P}. Like stellar mass density, the star formation rate has been shown to increase  with increasing redshift, up to an order of magnitude between $z=0.01$ and $z=1$ peaking between $z=2$ and $3$ \citep{2006ApJ...651..142H,2004ApJ...615..209H,2004AJ....128.2652G,2002AJ....124.1258W,2000AAS...196.0608H,1999ApJ...517..148F}. Though stellar mass density and star formation rate density have similar trends, the predicted stellar mass density from instantaneous star formation rate density measurements is higher than the observed stellar mass density by approximately $60\%$ \citep{2014ARA&A..52..415M}. Studies of molecular gas, the material from which stars form, have been biased towards gas rich, actively star forming galaxies, but indicate that gas mass fraction increases over redshift, and peaks at $z\sim2$ similar to the star formation rate \citep{2013ARA&A..51..105C}. Neutral atomic hydrogen (HI) in galaxies is the ultimate fuel for star formation, but the HI content of galaxies has only been measured in Damped Lyman-$\alpha$ systems beyond z$=2$ \citep{2012A&A...547L...1N}, and indirectly between $z\sim0.25$ and $z=2$. \citet{2011MNRAS.415...61S} found little correlation between M$_{HI}$ and M$_{H_{2}}$ in massive galaxies in the CO Legacy Database for GALEX Arecibo SDSS Survey. \citet{2015A&A...582A..78M} find that galaxies hosting long gamma-ray bursts are deficient in molecular gas but abundant in HI suggesting that at least the initial burst of star formation could come directly from the atomic gas. So though we may have a glimpse at how molecular gas changes as a function of redshift we are still unsure how the atomic gas, still an important element of star formation, changes as a function of redshift between $z=0$ and $z=1$.   

To date, atomic gas between $z=0$ and $z=0.2$ has been studied by measuring the $21$ cm HI emission. Surveys such as ALFALFA and HIPASS have detected large samples of galaxies and measured their HI content out to $z\sim0.08$ \citep{2005AJ....130.2598G,2003AJ....125.2842Z}, but until recently little was known about HI $21$ cm emission beyond $z=0.1$. \citet{2001Sci...293.1800Z} used the Westerbork Synthesis Radio Telescope (WSRT) to make the first detection of HI at z$>0.1$, z=$0.1766$, finding $M_{HI}=(6.0\pm0.8)\times10^{9}$ for a galaxy in the cluster Abell 2218.  \citet{,2008ApJ...685L..13C} and \citet{2010arXiv1009.0279V} detected HI in $\sim180$ galaxies between z=$0.16-0.25$ down to masses of M$_{HI}=3\times10^{10}$ to $2\times10^{9} M_{\odot}$ respectively with the Arecibo telescope and the Westerbork Synthesis Radio Telescope (WSRT) respectively. 

Indirect detections of HI have been made to $z=0.8$ using observing techniques such as stacking \citep{2009MNRAS.399.1447L} and intensity mapping \citep{2010Natur.466..463C}. \citet{2009MNRAS.399.1447L} found average HI mass of ($6.6\pm3.5$)$\times10^{9} M_{\odot}$ per galaxy in Abell 370, a cluster at $z=0.37$. Numerous groups have used the 100 m Green Bank Telescope\footnote{The National Radio Astronomy Observatory (NRAO) is a facility of the National Science Foundation operated under cooperative agreement by Associated Universities, Inc.}  (GBT), the only telescope with a cooled receiver that can detect HI at redshift $z\ge0.45$ with a reasonable integration time, to create an HI intensity map \citep{2010Natur.466..463C,2013ApJ...763L..20M,2013MNRAS.434L..46S}. After cross-correlating the GBT data with optical data from the WiggleZ Dark Energy Survey, \citet{2013ApJ...763L..20M} made a $7.4\sigma$ detection of HI density, $\Omega_{HI}=(0.4\pm0.05(stat.)\pm0.04(sys.))\times10^{-3}\times(1/rb)$, where r, the stochasticity, and b, the bias, are not well constrained.

Direct detections beyond $z=0.25$ are difficult due to the weakness of the 21 cm HI line, the limited sensitivity and frequency coverage of present-day radio telescopes, and the many sources of radio frequency interference (RFI) in the frequency bands that cover redshifted HI emission. After successful pilot observations \citep{2013ApJ...770L..29F}, the COSMOS HI Large Extragalactic Survey (CHILES) is currently using the recently upgraded Karl G. Jansky Very Large Array to search for $21$ cm emission in individual galaxies in the well observed COSMOS field out to $z=0.45$.  Telescopes designed specifically to carry out intensity mapping surveys like the Canadian Hydrogen Intensity Mapping Experiment (CHIME) \citep{2013MNRAS.434.1239B}, Baryon acoustic oscillations In Neutral Gas Observations (BINGO) \citep{2014SPIE.9145E..22B}, and Tianlai \citep{2015IAUGA..2252187C} will greatly improve measurements of $\Omega_{HI}$. Planned deep field surveys with telescopes such as ASKAP, MeerKAT and the Square Kilometer Array (SKA) \citep{Panorama,2012IAUS..284..496H,2015aska.confE.167S} will have the sensitivity and frequency coverage required to directly detect HI emission in individual galaxies beyond $z=0.45$. 

Until these telescopes are completed, direct detections of magnified HI 21 cm emission from strongly lensed sources can be made to $z\sim1$ using current telescopes \citep{2015MNRAS.452L..49D}. This technique has been used to detect magnified CO emission in many strongly lensed galaxies, the first being a galaxy at $z=2.2867$ by \citet{ 1991AJ....102.1956B}, and more recently a survey detecting emission from sources beyond $z=4$ \citep{2013Natur.495..344V}. Probing lensed sources for HI emission should also be possible, but careful concern is required when selecting these sources to ensure they are sufficiently magnified and the observed frequencies are not saturated with RFI. In this study we report on our observations of three gravitationally lensed galaxies behind the galaxy cluster Abell 773 \citep{2005ApJ...627...32S}, chosen because they had known redshifts and were likely to be highly magnified.

Throughout this paper we assume $H_{o}=69.7 km s^{-1} Mpc^{-1}, \Omega_{M}=0.282$, and $\Omega_{\Lambda}=0.718$ \citep{2013ApJS..208...19H} and use the cosmology calculator from \citet{2006PASP..118.1711W}\footnote{\url{http://www.astro.ucla.edu/~wright/CosmoCalc.html}} to calculate distances. In \S 2 we explain our observations, data reduction method, and the flagging method we used to remove RFI. In $\S 3$, we present the results. In $\S 4$, we discuss our results, the effectiveness of our flagging, and mass estimates based on galaxy magnitudes. In $\S 5$ we go through our conclusion and briefly discuss the application of our technique to additional targets.
\section{Observations and Analysis}
\label{obs}
\subsection{Sources}
\label{sources}
\begin{figure}[h!]
\figurenum{1}
\label{fig:labelledcluster}
\centering
\epsscale{1.2}
\plotone{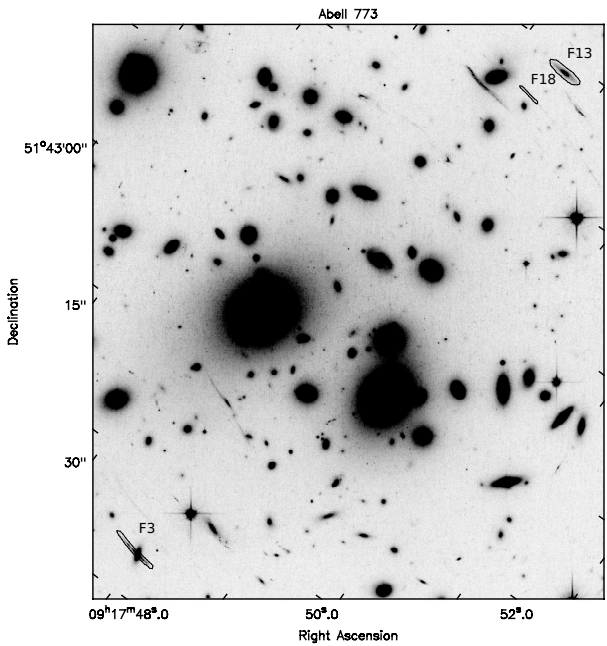}
\caption{The above figure shows an HST F814W image of the galaxy cluster Abell 773. The lensed galaxies F3, F13, and F18 are labelled and outlined. The positions for each source can be found in \citet{2005ApJ...627...32S}. The lensed galaxies should clearly fall within the 13' GBT beam.}
\end{figure}

Three sources were selected, F$3$, F$13$ and F$18$, from a list gravitationally lensed galaxies behind various massive clusters observed with the Hubble Space Telescope \citep{2005ApJ...627...32S}. These observations were taken using the Wide Field Planetary Camera 2 (WFPC2) on the \textit{Hubble Space Telescope} with the F702W filter. \citet{2005ApJ...627...32S} lists the magnitude of  F$3$ as $21.21\pm0.02$, F$13$ as $21.52\pm0.03$ and F$18$ as $23.39\pm0.11$.  F$3$ and F$13$ both lie at z$=0.398$ and F$18$ lies at z$=0.487$ \citep{2010MNRAS.404..325R,2005ApJ...627...32S}. The sources are labelled in Figure \ref{fig:labelledcluster}, and shows the they all fall within the $\sim13'$ GBT beam. 

We originally targeted these sources because they fell within $0.25\leq~z\leq1.0$ and had a high length-to-width ratio which suggested a higher magnification of $\sim10$. After our observations we obtained a lens model from J. Richard (2009, private communication) which yielded magnifications of $1.7\pm0.1$ for F3, $2.0\pm0.2$ for F13, and $2.7\pm0.3$ for F18 

\begin{table*}
\title{Summary of Observations}
\centering
\begin{tabular}{lccc}
\hline
\hline
& z=$0.398$ ($\nu=1016 MHz$) & z=$0.487$ ($\nu=955.5 MHz$) \\
\hline
Pointing Coordinates (J2000) &\multicolumn{2}{c}{ $09^{h}17^{m}54^{s}.62 , 51\degr43\arcmin44\arcsec.6$}\\
Observing Time (hr) & \multicolumn{2}{c}{77.6} \\
Effective Integration Time (hr)& $17.2$ & $18.1$\\
Bandwidth (MHz) & \multicolumn{2}{c}{10}\\
Frequency Resolution (kHz) & \multicolumn{2}{c}{39.1}\\
Velocity Resolution (km s$^{-1}$) & 11.5 & 12.3\\
System Temperature (K)&26.2&23.7\\
T$_{cal}$ (K)&$1.61$&$1.58$\\
$\%$ Flagged&6&0.85\\
\hline
\hline
\end{tabular}
\caption{Observing time is the total telescope time. Effective integration time is per polarization. Frequency and velocity resolutions listed are the final values after smoothing. T$_{cal}$ is the temperature injected by the noise diode used to compare what is measured by the telescope to a known value for calibration. The percentage of data flagged includes integrations that were dropped because of high noise, and the percentage of data flagged in Fourier space.}
\label{tab:tab1}
\end{table*}
\subsection{Observations}
Observations of the three galaxies were carried out, at night in order to minimize RFI, over seven years using the GBT which has a gain of $2$ K $Jy^{-1}$. Observations occurred $2007$ January $21$ and $26$, over eight sessions from $2008$, January, $16-24$, and over four sessions from $2014$, February, $2-6$. We used the Prime Focus $2$ receiver ($901-1230$ MHz), and the Spectral Processor backend with two polarizations and over two frequency bands for all observations. The first band was centered at $955$  MHz ($z=0.487$) and the second at $1016$ MHz (z=$0.398$), each with $10$ MHz bandwidth (${\Delta}v=3140$ km s$^{-1}$ at $955$ MHz and ${\Delta}v=2952$ km s$^{-1}$ at $1016$ MHz). In $2007$ and $2014$ our data had $512$ channels per frequency band for a frequency resolution of $19.5$ KHz and a velocity resolution of $6.2$ km s$^{-1}$ at $z=0.487$ and $5.8$ km s$^{-1}$ at $z=0.398$. In $2008$ the configuration changed to $256$ channels per frequency band with the final velocity resolution listed in Table \ref{tab:tab1}. The beam size for the PF2 receiver is $\sim13.6\arcmin$ at $955$ MHz and $\sim12.8\arcmin$ at $1018$ MHz; the three sources we observed were within $1\arcmin$ of our pointing direction so that all of the emission falls within the beam.

The GBT is located in the National Radio Quiet Zone which greatly reduces most, but not all, RFI from terrestrial sources. These sources of RFI cause both wide and narrow band interference with a wide range of flux density and will be further discussed in $\S$ 2.4. The Spectral Processor is ideally designed to handle the wide range of flux densities with a dynamic range of 45 dB.

Standard position switching was used for the observations, with a short integration time of two seconds to reduce the number of spectra affected by wideband RFI. In $2007$ we spent two minutes on source and two minutes off source, and switched to five minutes on source and five minutes off source for the $2008$ and $2014$ observations to reduce overhead. The off source pointing was carried out 2 minutes of right ascension, or 30 arcminutes, ahead of the source for the first year and 5 minutes of right ascension ahead of the source for the rest of the observations so that the same azimuth and zenith angle were tracked as the on source observations. Our total observing time was $77.6$ hours, giving an integration time of$17.2$ hours at $z=0.398$ and $18.1$ hours at $z=0.487$.

\subsection{Data Reduction}
Reduction of the data was done using GBTIDL\footnote{GBTIDL (\url{http://gbtidl.nrao.edu/}) is the data reduction package produced by NRAO and written in the IDL language for the reduction of GBT data.}. For each night, spectral window, on-off scan pair, polarization, and integration we used the \textit{getps} procedure which calibrates a total power, position switched scan pair as defined in Equation \ref{eq:equation 1} where T$_{sys}$ is listed in Table \ref{tab:tab1}, the gain for the GBT is 2 Jy K$^{-1}$ and the atmospheric opacity correction is $\tau=0.01$.
\begin{center}
\begin{equation}
S=\frac{On-Off}{Off}\times T_{sys}\times~Gain\times~e^{\tau}
\label{eq:equation 1}
\end{equation}
\end{center}
Next, a fifth order polynomial baseline, the lowest order polynomial that produces flat baselines, was fit across the frequencies $950.24$-$953.65$ MHz, $954.15-957.25$ MHz and $957.95-960$ MHz at $z=0.487$ and $1011.25-1011.75$ MHz and $1012.5-1021$ MHz at $z=0.398$. This fit was applied to every integration whether wideband RFI was present or not. The fifth order polynomial applied across our 10 MHz bandwidth removes variations in the bandpass of roughly 2 MHz and should not affect a typical galaxy with line width of $200 km s^{-1}$ \citep{2011ApJ...739...38P} which would have a frequency width of $\lesssim677$ kHz. We developed an automated flagging routine, described in more detail in Section 2.4, which we then used to flag all RFI. A spectrum for each night and spectral window was created by accumulating integrations from all on-off scan pairs and both polarizations. The effective integration time at $1016$ MHz is much lower, and percentage of data flagged much higher, because the data for one polarization on night two had abnormally high noise, and was flagged completely. There is no obvious reason for why the noise was higher in that one polarization. Next, we  used a Gaussian curve to smooth over the first two nights and the last four nights to change the frequency resolution from 19.5 kHz per channel (a velocity resolution of $6.2$ km s$^{-1}$ at $z=0.487$ and $5.8$ km s$^{-1}$ at $z=0.398$) to 39.1 kHz per channel (a velocity resolution of $12.3$ km s$^{-1}$ at $z=0.487$ and $11.5$ km s$^{-1}$ at $z=0.398$) to improve the noise and match the other eight nights. A fifth order polynomial baseline was fit to the final spectrum for each night and then we accumulated and averaged all nights to get the final spectrum.

\subsection{RFI and Flagging Routine}

The observed frequency bands contain a large amount of RFI which we believe is caused by airplanes distance measuring equipment (DME) radar. The ground to air portion of the radar transmits between $962$ and $1024$ MHz. \citet{2005AJ....129.2940F} describes DME as a pair of strong pulses between an airplane and a ground station sent at 24 to 30 pairs a second, meaning that each pulse is much shorter than the two second integration time. Our two second integration time is a limit of the Spectral Processor, but future observations will be able to take advantage of the increased dynamic range and processing power of the VErsatile GBT Astronomical Spectrometer (VEGAS) allowing for increased bandwidth and spectral resolution, and shorter integration time. The wideband interference could come from a strong, intermittent DME signal outside of the observing band, appearing consecutively in up to 77 integrations. The time the RFI was visible in the data ranged from 2 seconds to 144 seconds, but was most frequently visible for $\sim 4$ seconds at a time. The wideband RFI has a characteristic width of 1 MHz which is still larger than the aforementioned 677 kHz frequency expected for the signal.
 
Both wideband and narrow band interference were removed using a custom sigma clipping method \citep{1977ApJ...214..347Y}, measuring the standard deviation of the spectrum and removing points above or below some multiple of that value. Channels around the clipped point were then blanked to remove the whole spike. This was done on the frequency spectrum (the frequency domain) to remove narrow band interference, and on its Fourier transform (the Fourier domain) to remove wideband interference.

We only wanted to flag spectra in the Fourier domain when wideband RFI was present in order to avoid unnecessarily removing data. The data containing wideband interference often had tall spikes in the Fourier transform, and we used this to search for wideband interference. We also found the data that contained narrow band interference had tall spikes in Fourier space. From data analysis we saw that the narrow band interference occurred most frequently at $954$ and $958$ MHz at $z=0.487$ and $1012$ MHz and $1018$ MHz at $z=0.398$. Before we carried out our preliminary Fourier transform to search for wideband RFI, we quickly interpolated over those frequencies. This way we were able to test for wideband interference using the Fourier domain while avoiding confusion with narrow band interference. If a peak in Fourier space was measured over our threshold, a flux density greater than $0.015$ Jy or less than -$0.015$ Jy, we assume that integration contains wideband interference. 

We then started over with the original spectrum and continued our clipping routine. The values were selected after measuring the maximum and minimum values in the Fourier domain of many spectra both containing and lacking wideband RFI. The unaltered spectra were Fourier transformed and clipped, setting channels larger than $4.3\sigma$ to zero and doing the same for four channels on either side when the band had 512 channels and two channels on either side when the band had 256 channels to ensure the spike was removed. The value of $4.3\sigma$ and the number of channels flagged on either side of the spike were both chosen after testing various combinations to determine the lowest combination that removed all wideband interference. After the spikes in the Fourier domain were removed, the spectrum was inverse Fourier transformed, and the wideband interference was no longer present in the frequency domain. We continued in the frequency domain, measuring the standard deviation across the central 2 MHz, blanking channels larger than 3.5$\sigma$, and removing seven channels on either side when the band had 512 channels, and four channels on either side when the band had 256 channels, eliminating most of the narrowband interference in the frequency domain. Because the narrow band RFI is ever present at $954$ and $957.5$ MHz we could not remove all of it from the final spectrum. Again the value for sigma clipping and width were determined by testing multiple scans to find the lowest combination of values to remove as much of the spike as possible without removing signal.

\begin{figure}[htb!]
\figurenum{2}
\epsscale{1.2}
\label{fig:Test}
\centering
\plotone{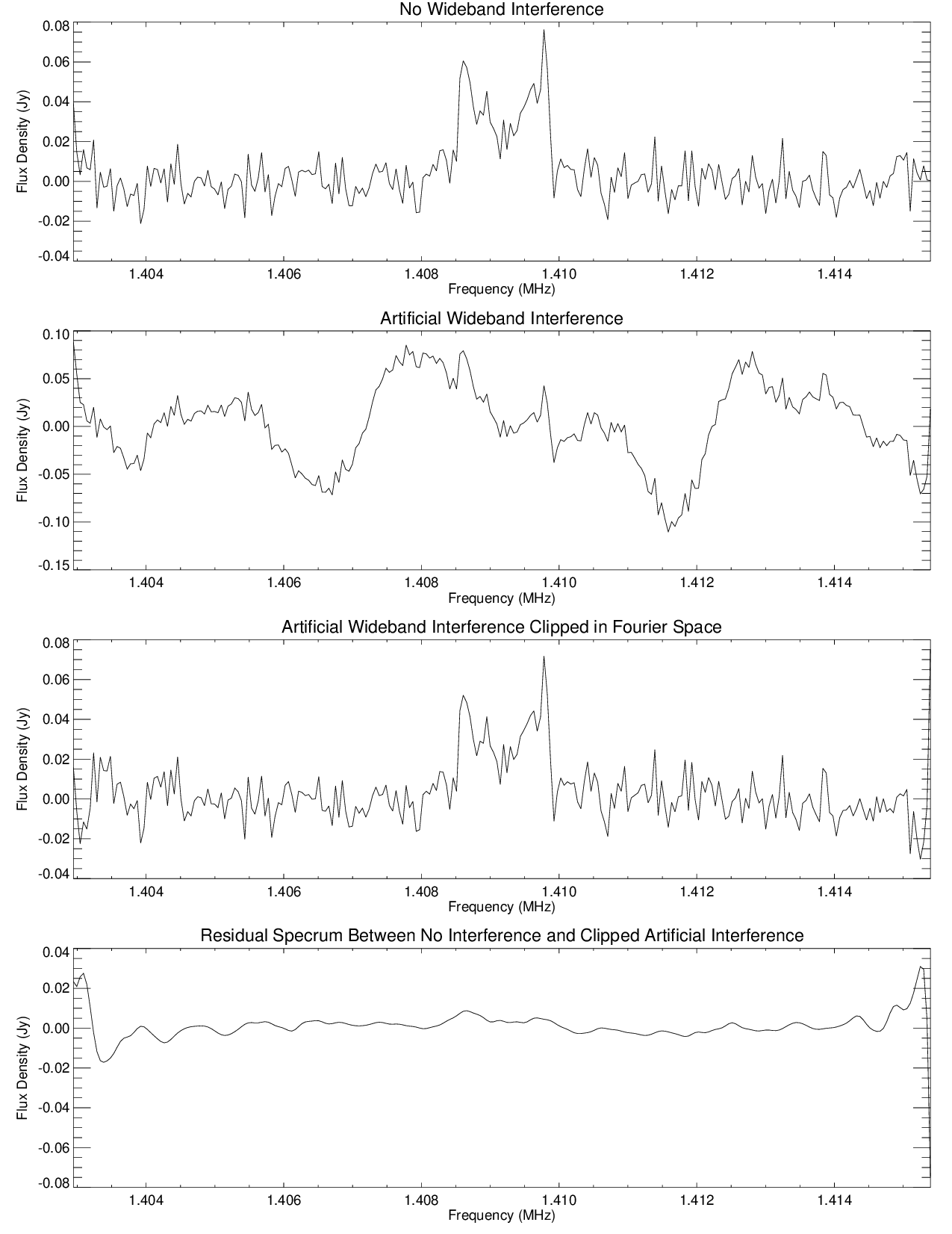}
\caption{The above figures show the spectrum we used to test our flagging routine. The first plot shows the final spectrum without adding wideband interference. The second plot shows the final spectrum after adding wideband interference to $10\%$ of the integrations. The third plot shows the final spectrum after adding wideband interference and then applying our flagging routine. The fourth plot shows the difference between the spectrum in which no interference was added, and the spectrum which was flagged in Fourier space.}
\end{figure}

A test was developed to check the effectiveness of our flagging routine. We used a GBT observation of the galaxy NGC $5375$ in which the signal from the galaxy is not visible in a single integration, but becomes visible when many integrations are averaged together. NGC $5375$ is at much lower redshift and has a velocity width of 280 km s$^{-1}$, so its frequency width, 1.4 MHz, is 2.1 times larger than the assumed frequency width of the high redshift sources, 677 kHz. To make sure the HI signal is not visible in a single integration we artificially increased the noise in each channel in each integration by adding a random number to the measured flux density in each channel. The signal to noise ratio in each integration becomes 0.66 after adding this artificial noise, making the source undetectable in a single integration. Then we introduced wideband interference, extracted from our original dataset by fitting a high order polynomial, to approximately 10$\%$ of the integrations in the test dataset at random. It is important to note that we only flagged about 190 channels, much less than $1\%$ of the data. Next we flagged the data in the Fourier domain for the integrations in which we introduced wideband interference, zeroing any values higher than $4.3$ times the noise like we did to the sources behind Abell $773$, and did nothing to the others. We averaged all of the integrations to create a final average spectrum, and compared them to the unaltered final spectrum. The results in Figure \ref{fig:Test} show that the HI signal looks the same when there is no wideband interference, and when the artificial wideband interference is removed using our flagging routine. The spectrum without wideband interference added had an RMS of $0.0093$ Jy with an integrated signal to noise ratio of 8.2 and the spectrum with wideband interference added and then removed had an RMS of $0.0095$ Jy with an integrated signal to noise ratio of 7.6. The RMS of the residual spectrum is $0.002$ Jy. The signal to noise ratio and frequency width of the test source are much larger than those expected from our data, so the test represents an extreme and the flagging routine should have a smaller effect on our data. 
\section{Results}
\label{sec:results}
The final spectra are shown in Figure 2. The spectra have an RMS of $ 211~\mu$Jy at $z=0.487$ and $204 ~ \mu$Jy at $z=0.398$. For comparison, the theoretical value for the noise is $183~\mu$Jy and $171~\mu$Jy respectively. This is calculated from
\begin{center}
\begin{equation}
\sigma=\frac{T_{sys}}{G~\sqrt{N_{pol}~\Delta\nu~t_{eff}}}
\label{eq:eq2}
\end{equation}
\end{center}
where $T_{sys}$ is the system temperature, $23.7$ K at $\nu=1016$ MHZ and 26.2 K at $\nu=955$ MHZ, $G=2 K Jy^{-1}$ is the gain, $N_{pol}=2$ is the number of polarizations, $\Delta\nu=39.1$ KHz is the frequency width per channel, and $t_{eff}$ is the effective integration time listed in Table \ref{tab:tab1}. The corresponding M$_{HI}$ detection limit is
\begin{center}
\begin{equation}
\sigma_{MHI}=\frac{2.36\times10^{5}D_{L}^{2}\sigma_{s}dv\sqrt{N}\sigma_{SNR}}{(1+z)\mu}
\label{eq:eq3}
\end{equation}
\end{center}
where $\sigma_{MHI}$ is the upper limit of the neutral hydrogen mass, $D_{L}$ is the luminosity distance to the object, $2.18$ Gpc at $z=0.398$ and $2.77$ Gpc at $z=0.487$ , $\sigma_{s}$ is the RMS per channel in the spectrum in units of Jy, dv is the channel width in km s$^{-1}$, N is the number of channels the galaxy would span, $\sigma_{SNR}$ is the signal to noise ratio, z is redshift, and $\mu$ is the magnification.  To set a mass limit we need to select a value for N. We use the mode of the line width, $\sim200$ km s$^{-1}$, from $10744$ galaxies in the Arecibo Legacy Fast ALFA (ALFALFA) survey \citep{2011ApJ...739...38P} divided by a velocity resolution of $12.3$ km s$^{-1}$ at $z=0.487$ and $11.5$ km s$^{-1}$ at $z=0.398$ to find $N=16$ and $17$ respectively. Using the above parameters, we calculate the $3$$\sigma$ detection limit to be M$_{HI}=1.50\times$ $10^{10}$ $M_{\odot}$ and M$_{HI}=6.36\times10^{9}$ $M_{\odot}$ respectively.

\begin{figure}[h!]
\label{fig:finalspectra}
\figurenum{3}
\epsscale{1.3}
\plotone{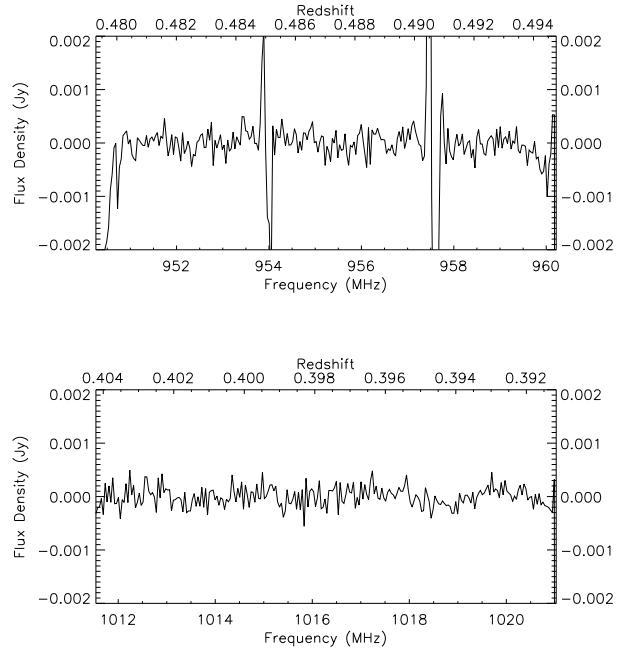}
\caption{This is the final combined spectrum for our data at both $z=0.398$ and $z=0.487$. The RFI present at $954$ MHz and $957.5$ MHz persisted no matter how we changed our flagging routine.}
\end{figure}

\section{Discussion}
\subsection{Effectiveness of Flagging}
The flagging procedure we used was possible only because the galaxy is not visible in any single integration and there was no danger of clipping or altering our signal. As mentioned in $\S 2.2$, we measured the standard deviation of the central $2$ MHz ($628$ km s$^{-1}$ at $955$ MHz and $590$ km s$^{-1}$ at $1016$ MHz) of the spectrum in both bands and used that value for sigma-clipping. We did this because that region of the spectrum was generally devoid of narrow-band RFI. In the central $2$ MHz very few integrations were clipped in the frequency domain, so besides the one polarization in night two that was discarded due to high noise, most of the information that was lost came from clipping in the Fourier domain. Only $6.53\%$ of the data was flagged at $1016$ MHz and $1.6\%$ was flagged at $955$ MHz. If we ignored integrations with wideband RFI instead of flagging in Fourier space, we would have removed $\sim14\%$ of the data at $955$ MHz and $\sim10\%$ of the data at $1016$ MHz. While this is a small effect on the final noise of the spectrum, it also yields cleaner baselines.

\begin{figure*}[htb!]
\figurenum{4}
\label{fig:Comparison}
\epsscale{1.2}
\plotone{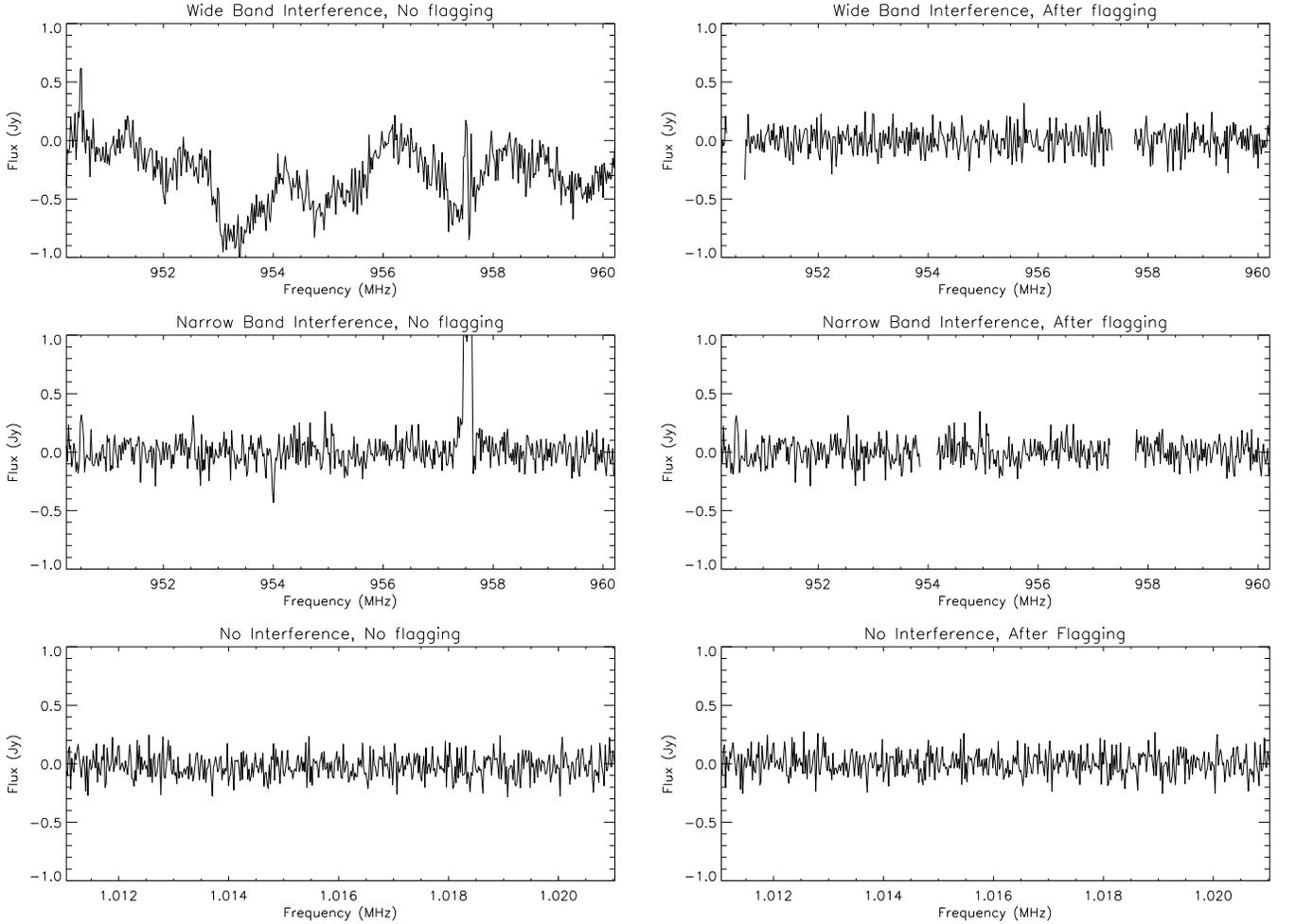}
\caption{The above figures show the spectrum before and after the clipping procedure. The first example shows the clipping of a single integration that showed wideband and narrowband interference. You can see that the wave structure was removed and two spikes were removed. The second example shows clipping of only narrow band interference. The third example shows a spectrum that doesn't appear to have any interference. We measured the standard deviation before and after clipping and found that the spectrum was unaffected. Standard deviation was $\sigma=0.102$ Jy for each spectrum}
\end{figure*}

In Figure \ref{fig:Comparison} we show three examples of spectra before and after flagging. The top panel shows a spectrum with wideband RFI before and after flagging. In the first spectrum, our routine was effective at removing the wideband RFI and we were also able to remove narrowband spikes. The middle panel has a spectrum showing only narrow band RFI. The spectrum was not changed in Fourier space and the narrow band RFI was removed with sigma clipping. The bottom example in Figure \ref{fig:Comparison} shows flagging done when the spectrum does not appear to contain any RFI. The spectrum before and after the flagging routine remained exactly the same.

\begin{figure}[htb!]
\figurenum{5}
\label{fig:FlagStats}
\epsscale{1.2}
\plotone{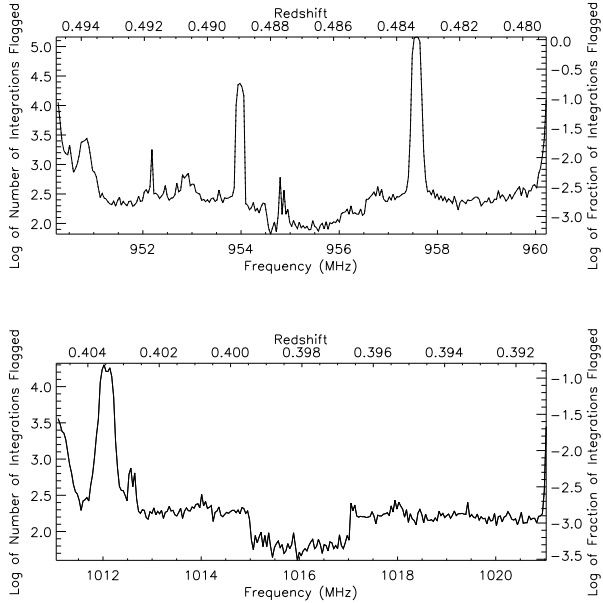}
\caption{The above figures shows how often each frequency bin was clipped. The dip in the middle of each band is due to setting the standard deviation based on those frequencies. Since our statistics were based off that region, fewer data points were outside 3.5 times the standard deviation.}
\end{figure}

Figure \ref{fig:FlagStats} shows the channels that were clipped most often in the frequency domain. In the band centered at $955$ MHz (z=$0.398$) narrow band interference was frequently present near $954$ MHz and $958$ MHz. After changing our widening parameters we were still unable to remove it entirely from our final spectrum and it is still visible in Figure \ref{fig:finalspectra}. We removed many channels on the edge and around 1012 MHz in the band centered at $1016$ MHz (z=$0.487$). A large percentage of the data is flagged because of narrow band RFI outside the central region. The dips in the middle of each plot in Figure \ref{fig:FlagStats} correspond to the areas where we measured the standard deviation and retained most of the data.

\subsection{Mass Limits}
 We were able to use the scaling relation between R band absolute magnitude and HI mass described by  \cite{2014MNRAS.444..667D} derived from the HI Parkes All Sky Survey (HIPASS), to estimate the HI mass for each galaxy. A galaxy type for these sources was not available so we assumed all three were Sbc spirals, the best case scenario for detection. The estimated masses are $M_{HI_{F3}}=(2.2$$\pm1.3)\times 10^{9}$, $M_{HI_{F13}}=(1.5$$\pm.8)\times 10^{9}$, and $M_{HI_{F18}}=(4.7$ $\pm2.3)\times 10^{8}$. The calculated average mass weighted by the magnification for F$3$ and F$13$ is $M_{HI_{F3+F13}}=(1.8$$\pm0.7)\times 10^{9}$. We obtained the local unmagnified R band magnitude by transforming the magnified, redshifted F702W magnitude using the k-corrections and color relations from \citet{1995PASP..107..945F}. F$3$ and F$13$ lie at the same redshift so we add the mass of the sources together weighted by their magnification for a total estimated mass detectable at $z=0.398$. These mass estimates are $\sim30$ and $3.7$ times smaller than the detection limit for F$18$ and both F$3$ and F$13$ respectively. To bring the noise down to the level required to detect emission from F$3$ and F$13$ we require approximately $200$ extra hours of integration time, or $800$ hours of observation time. 

\begin{figure}[h!]
\figurenum{6}
\label{fig:MagPlot}
\epsscale{1.15}
\plotone{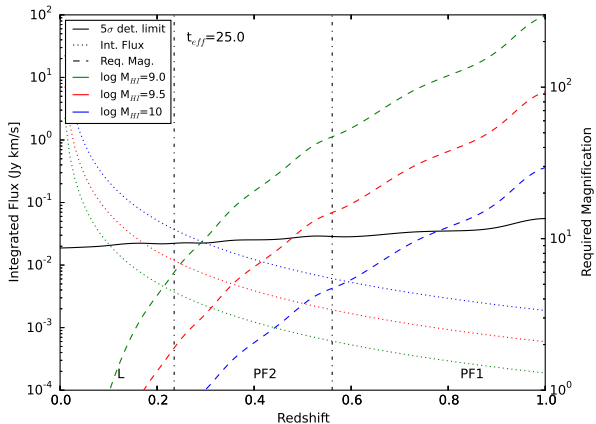}
\caption{This figure shows the magnification necessary to detect a galaxy of M$_{HI}=10^{9}$ M$_{\odot}$ (green), $10^{9.5}$ M$_{\odot}$ (red), $10^{10}$ M$_{\odot}$ (blue) at a given redshift using the GBT with a $25$ hour effective integration time. The dotted line represents the integrated flux for each mass, the solid line indicates the $5\sigma$ detection limit with $25$ hour effective integration time, and the dashed line indicates the magnification required to bring the integrated flux of the source up to the detection limit. The detection limit is calculated using the T$_{sys}$ values of the GBT and fluctuates in accordance.}
\end{figure}

Our detection limit would have been much lower if the length-to-width ratio had been a more accurate predictor of magnification, but future studies can target objects with known magnifications from more accurate lens models. We can use Figure \ref{fig:MagPlot} to determine the magnification required to detect sources of various mass with $25$ hour integration time. For example, we should be able to detect a galaxy with M$_{HI}=3.16\times10^{9}$ out to z=$0.725$ as long as it has a $\mu>30$. A strongly lensed arc behind the cluster Abell 370 at z=$0.725$ has been mentioned in \citet{2010MNRAS.404..325R}. It appears to be an SBc type galaxy, which typically have higher HI mass \citep{1994ARA&A..32..115R}, with a total magnification of 32, and should be detectable with the GBT within 100 hours of observation time.   

\section{Conclusions and Future Work}
Gravitational lensing has been used to measure emission from molecular gas and stars at higher redshift in the past and we suggest it can be used to measure HI at high redshift with current telescopes. The lower frequencies of redshifted HI have a denser RFI environment and, as our observations show, can lead to over $10\%$ of the data being discarded due to wideband RFI. To recover as much information as possible from our data we created a custom sigma clipping routine that removes wideband and narrowband interference. The narrowband interference is removed using sigma clipping in the frequency domain, and wideband interference is removed using sigma clipping in the Fourier domain.  When we flag that data in the Fourier domain the information near the source is preserved and the wideband RFI removed, reducing the noise and flattening the baseline. Testing shows the routine is effective when the narrowband interference does not cover your signal, and when  $\lesssim10\%$ of the integrations have wideband interference. We used this flagging technique on spectra from three sources behind the galaxy cluster Abell 773. Flagging spectrum with wideband interference in the Fourier domain rather than discarding them reduced the final noise in the spectrum by about 25\%.  The final noise per channel of our spectrum was $211 \mu Jy$ at $z=0.487$ and $204 \mu Jy$ at $z=0.398$. This flagging method can be combined with the magnification provided by gravitational lensing to detect HI in individual galaxies out to and beyond a redshift of $z=0.5$

We have observed three sources for an effective integration time of $18.1$ hrs at $z=0.487$ and $17.2$ hrs at $z=0.398$. The sources have low magnification, and were not detected, but we were able to set a 3$\sigma$ detection limit on their masses of M$_{HI}=1.35\times$ $10^{10}$ $M_{\odot}$ and M$_{HI}=6.36\times10^{9}$ $M_{\odot}$ respectively which is higher than their estimated masses of $4.7\times 10^{8}$ and $1.8\times10^{9}$ $M_{\odot}$ respectively. In order to detect these sources we require 235 additional hours of observation time. Adding more observation time now seems costly, so we must identify other sources to observe that should be detectable based on their magnification.

We have used the criteria from Figure \ref{fig:MagPlot} to find other sources that are detectable with the GBT. Sources beyond $z=1$ would require unrealistically large magnifications, effective  integration times, or HI masses to be detectable so we have identified 28 lenses below that redshift that can be targeted for similar observations. We have already begun observations on two sources behind the cluster Abell $370$, one with a magnification of $\sim32$ where an HI detection should be possible in an effective integration time of 25 hours.


\begin{thebibliography}{99}
\bibitem[Battye et al.(2013)]{2013MNRAS.434.1239B} Battye, R.~A., Browne, I.~W.~A., Dickinson, C., et al.\ 2013, \mnras, 434, 1239 
\bibitem[Bandura et al.(2014)]{2014SPIE.9145E..22B} Bandura, K., Addison, G.~E., Amiri, M., et al.\ 2014, \procspie, 9145, 914522 
\bibitem[Brown \& Vanden Bout(1991)]{1991AJ....102.1956B} Brown, R.~L., \& Vanden Bout, P.~A.\ 1991, \aj, 102, 1956
\bibitem[Carilli \& Walter(2013)]{2013ARA&A..51..105C} Carilli, C.~L., \& Walter, F.\ 2013, \araa, 51, 105 
\bibitem[Catinella et al.(2008)]{2008ApJ...685L..13C} Catinella, B., Haynes, M.~P., Giovanelli, R., Gardner, J.~P., \& Connolly, A.~J.\ 2008, \apjl, 685, L13
\bibitem[Chang et al.(2010)]{2010Natur.466..463C} Chang, T.-C., Pen, U.-L., Bandura, K., \& Peterson, J.~B.\ 2010, \nat, 466, 463
\bibitem[Chen(2015)]{2015IAUGA..2252187C} Chen, X.\ 2015, IAU General Assembly, 22, 2252187
\bibitem[D{\'e}nes et al.(2014)]{2014MNRAS.444..667D} D{\'e}nes, H., Kilborn, V.~A., \& Koribalski, B.~S.\ 2014, \mnras, 444, 667
\bibitem[Deane et al.(2015)]{2015MNRAS.452L..49D} Deane, R.~P., Obreschkow, D., \& Heywood, I.\ 2015, \mnras, 452, L49 
\bibitem[Dickinson et al.(2003)]{2003ApJ...587...25D} Dickinson, M., Papovich, C., Ferguson, H.~C., \& Budav{\'a}ri, T.\ 2003, \apj, 587, 25
\bibitem[Fern{\'a}ndez et al.(2013)]{2013ApJ...770L..29F} Fern{\'a}ndez, X., van Gorkom, J.~H., Hess, K.~M., et al.\ 2013, \apjl, 770, L29
\bibitem[Fisher et al.(2005)]{2005AJ....129.2940F} Fisher, J.~R., Zhang, Q., Zheng, Y., Wilson, S.~G., \& Bradley, R.~F.\ 2005, \aj, 129, 2940
\bibitem[Flores et al.(1999)]{1999ApJ...517..148F} Flores, H., Hammer, F., Thuan, T.~X., et al.\ 1999, \apj, 517, 148
\bibitem[Fontana et al.(2003)]{2003ApJ...594L...9F} Fontana, A., Donnarumma, I., Vanzella, E., et al.\ 2003, \apjl, 594, L9
\bibitem[Fukugita et al.(1995)]{1995PASP..107..945F} Fukugita, M., Shimasaku, K., \& Ichikawa, T.\ 1995, \pasp, 107, 945
\bibitem[Giovanelli et al.(2005)]{2005AJ....130.2598G} Giovanelli, R., Haynes, M.~P., Kent, B.~R., et al.\ 2005, \aj, 130, 2598 
\bibitem[Glazebrook et al.(2004)]{2004AJ....128.2652G} Glazebrook, K., Tober, J., Thomson, S., Bland-Hawthorn, J., \& Abraham, R.\ 2004, \aj, 128, 2652
\bibitem[Haarsma et al.(2000)]{2000AAS...196.0608H} Haarsma, D.~B., Partridge, R.~B., Windhorst, R.~A., \& Richards, E.~A.\ 2000, Bulletin of the American Astronomical Society, 32, 685
\bibitem[Hinshaw et al.(2013)]{2013ApJS..208...19H} Hinshaw, G., Larson, D., Komatsu, E., et al.\ 2013, \apjs, 208, 19 
\bibitem[Holwerda et al.(2012)]{2012IAUS..284..496H} Holwerda, B.~W., Blyth, S.-L., Baker, A.~J., \& Baker 2012, IAU Symposium, 284, 496
\bibitem[Hopkins \& Beacom(2006)]{2006ApJ...651..142H} Hopkins, A.~M., \& Beacom, J.~F.\ 2006, \apj, 651, 142 
\bibitem[Hopkins(2004)]{2004ApJ...615..209H} Hopkins, A.~M.\ 2004, \apj, 615, 209
\bibitem[Ilbert et al.(2013)]{2013A&A...556A..55I} Ilbert, O., McCracken, H.~J., Le F{\`e}vre, O., et al.\ 2013, \aap, 556, A55
\bibitem[Lah et al.(2009)]{2009MNRAS.399.1447L} Lah, P., Pracy, M.~B., Chengalur, J.~N., et al.\ 2009, \mnras, 399, 1447
\bibitem[Madau \& Dickinson(2014)]{2014ARA&A..52..415M} Madau, P., \& Dickinson, M.\ 2014, \araa, 52, 415 
\bibitem[Marchesini et al.(2009)]{2009ApJ...701.1765M} Marchesini, D., van Dokkum, P.~G., F{\"o}rster Schreiber, N.~M., et al.\ 2009, \apj, 701, 1765
\bibitem[Masui et al.(2013)]{2013ApJ...763L..20M} Masui, K.~W., Switzer, E.~R., Banavar, N., et al.\ 2013, \apjl, 763, L20 
\bibitem[Meyer(2009)]{Panorama} Meyer, M.\ $2009$, \ in Panoramic Radio Astronomy$:$ Wide-field $1-2$ GHz Research on Galaxy Evolution
\bibitem[Micha{\l}owski et al.(2015)]{2015A&A...582A..78M} Micha{\l}owski, M.~J., Gentile, G., Hjorth, J., et al.\ 2015, \aap, 582, A78 
\bibitem[Noterdaeme et al.(2012)]{2012A&A...547L...1N} Noterdaeme, P., Petitjean, P., Carithers, W.~C., et al.\ 2012, \aap, 547, LL1 
\bibitem[P{\'e}rez-Gonz{\'a}lez et al.(2008)]{2008ApJ...675..234P} P{\'e}rez-Gonz{\'a}lez, P.~G., Rieke, G.~H., Villar, V., et al.\ 2008, \apj, 675, 234
\bibitem[Papastergis et al.(2011)]{2011ApJ...739...38P} Papastergis, E., Martin, A.~M., Giovanelli, R., \& Haynes, M.~P.\ 2011, \apj, 739, 38
\bibitem[Richard et al.(2010)]{2010MNRAS.404..325R} Richard, J., Smith, G.~P., Kneib, J.-P., et al.\ 2010, \mnras, 404, 325
\bibitem[Roberts \& Haynes(1994)]{1994ARA&A..32..115R} Roberts, M.~S., \& Haynes, M.~P.\ 1994, \araa, 32, 115 
\bibitem[Rudnick et al.(2003)]{2003ApJ...599..847R} Rudnick, G., Rix, H.-W., Franx, M., et al.\ 2003, \apj, 599, 847
\bibitem[Saintonge et al.(2011)]{2011MNRAS.415...61S} Saintonge, A., Kauffmann, G., Wang, J., et al.\ 2011, \mnras, 415, 61
\bibitem[Sand et al.(2005)]{2005ApJ...627...32S} Sand, D.~J., Treu, T., Ellis, R.~S., \& Smith, G.~P.\ 2005, \apj, 627, 32
\bibitem[Staveley-Smith \& Oosterloo(2015)]{2015aska.confE.167S} Staveley-Smith, L., \& Oosterloo, T.\ 2015, Advancing Astrophysics with the Square Kilometre Array (AASKA14), 167 
\bibitem[Switzer et al.(2013)]{2013MNRAS.434L..46S} Switzer, E.~R., Masui, K.~W., Bandura, K., et al.\ 2013, \mnras, 434, L46 
\bibitem[Verheijen et al.(2010)]{2010arXiv1009.0279V} Verheijen, M., Deshev, B., van Gorkom, J., et al.\ 2010, arXiv:1009.0279
\bibitem[Vieira et al.(2013)]{2013Natur.495..344V} Vieira, J.~D., Marrone, D.~P., Chapman, S.~C., et al.\ 2013, \nat, 495, 344 
\bibitem[Wilson et al.(2002)]{2002AJ....124.1258W} Wilson, G., Cowie, L.~L., Barger, A.~J., \& Burke, D.~J.\ 2002, \aj, 124, 1258
\bibitem[Wright(2006)]{2006PASP..118.1711W} Wright, E.~L.\ 2006, \pasp, 118, 1711
\bibitem[Yahil \& Vidal(1977)]{1977ApJ...214..347Y} Yahil, A., \& Vidal, N.~V.\ 1977, \apj, 214, 347
\bibitem[Zwaan et al.(2001)]{2001Sci...293.1800Z} Zwaan, M.~A., van Dokkum, P.~G., \& Verheijen, M.~A.~W.\ 2001, Science, 293, 1800
\bibitem[Zwaan et al.(2003)]{2003AJ....125.2842Z} Zwaan, M.~A., Staveley-Smith, L., Koribalski, B.~S., et al.\ 2003, \aj, 125, 2842
\end{thebibliography}
\end{document}